\documentstyle[aps,psfig,preprint]{revtex}
\input epsf

\renewcommand{\thanks}{\footnote}
\def\tocite#1{$^{\hbox{\,-}}$\kern-.04em\cite{#1}}

\def\JLone<#1,#2>{#1}
\def\JLtwo<#1,#2,#3>{#2}
\def\JLyear<#1,#2,#3,#4>{#3}
\def\JLpage<#1,#2,#3,#4>{#4}
\def\JL#1{\JLone<#1>\ {\bf \JLtwo<#1>} (\JLyear<#1>), \JLpage<#1>}
\def\Jpage<#1,#2,#3>{#3}
\def\andvol#1{{\bf \JLone<#1>} (\JLtwo<#1>), \Jpage<#1>}
\def\PTP#1{Prog.\ Theor.\ Phys.\ \andvol{#1}}

\def\PR#1{Phys.\ Rev.\ \andvol{#1}}
\def\PRL#1{Phys.\ Rev.\ Lett.\ \andvol{#1}}
\def\PL#1{Phys.\ Lett.\ \andvol{#1}}
\def\NP#1{Nucl.\ Phys.\ \andvol{#1}}

  \def\gsl#1{\rlap{\slash}#1}
  \def\M{M}
  \def\tr{{\rm tr}}
  \def\vic{\sim}

  \def\vec#1{\mbox{\boldmath $#1$}}
  \def\gvec{\vec \gamma}

  \def\E{E_{\vec q}}
  \def\EP{E_{\vec q'}}

  \def\M{M}

  \def\w{\omega_0}
  \def\wp{\omega_+}
  \def\wm{\omega_-}

  \def\MB{M_B}
  \def\MBS{M_B^2}
  
  \def\MBQ{M_B^4}

\preprint{
KEK Preprint 98-92
}

\title{
QCD Sum Rules for Nucleon-Nucleon Interactions
}

\author{
Yoshihiko {\sc Kondo}\footnote{E-mail: kondo@tanashi.kek.jp} 
and Osamu {\sc Morimatsu$^{**}$}\footnote{E-mail: osamu.morimatsu@kek.jp}
}

\address{
$^{*)}$Kokugakuin University, Higashi, Shibuya, Tokyo 150-8440, Japan
\\
$^{**)}$Institute of Particle and Nuclear Studies, High Energy Accelerator Research Organization, Tanashi, 188-8501, Japan
}

\begin{document}

\maketitle

\abstract{
\noindent
The QCD sum rules for spin-dependent nucleon-nucleon interactions are formulated and their physical implications are studied.
The basic object of the study is the correlation function of the nucleon interpolating field, where the matrix element is taken with respect to the one-nucleon state.
By means of the dispersion relation, the correlation function in the deep Euclidean region, where it is expressed in terms of the nucleon matrix elements of the quark-gluon composite operators by using the operator product expansion (OPE), is related with its integral over the physical region.
The dispersion integral of the correlation function around the nucleon threshold is investigated in detail.
It turns out that the integral can be identified as a measure of the nucleon-nucleon interaction strength, which is proportional to the scattering length in the small scattering length limit and to one half of the effective range in the large scattering length limit.
New operators, such as $\bar q\gamma_\mu\gamma_5q$, $\bar q\gamma_5\sigma_{\mu\nu}q$, must be taken into account in the OPE of the correlation function. Therebehavior operators do not vanish when the matrix element is taken with respect to the spin-nonaveraged one-nucleon state.
The Wilson coefficients of such operators are calculated. 
The sum rules obtained in this manner relate the spin-dependent nucleon-nucleon interaction strengths with the spin-dependent nucleon matrix elements of the quark-gluon composite operators.
The sum rules imply that the interaction is stronger in the spin-triplet channel than in the spin-singlet channel, but that the spin-dependence of the nucleon-nucleon interactions is rather small.
In the spin-singlet channel the calculated strength is in qualitative agreement with the empirical strength, which is estimated by the empirical low energy scattering observables.
}

\section{Introduction}
One of the most important goals of investigating strong interaction physics is to understand the  of hadrons and hadronic interactions on the basis of quantum chromodynamics (QCD).
Shifman, Vainshtein and Zakharov proposed the method of the QCD sum rule, which provides us with a framework to investigate the properties of hadrons in a model-independent way.~\cite{rf:SVZ}
This method has been successfully applied to the study of the masses, decay constants, magnetic moments and other properties of various hadrons.~\cite{rf:RRY}

Recently, the present authors extended the QCD sum rule to the investigation of hadronic interactions.~\cite{rf:KM}
In Ref.~3) a nucleon-nucleon system was studied as a typical case.
The correlation function of the nucleon interpolating field, whose matrix element is taken with respect to the one-nucleon state and averaged over the nucleon spin, was considered.
It was noted that the correlation function has a second-order pole at the nucleon on-shell energy as a function of the energy associated with the interpolating field and that its coefficient is the T-matrix for the nucleon-nucleon ($NN$) scattering.
Assuming that the dispersion integral is dominated by the pole term, sum rules were derived which relate the spin-averaged $NN$ scattering lengths with the spin-averaged matrix elements of the quark-gluon operators with respect to the one-nucleon state.
The obtained $NN$ scattering lengths are of the order of several fm, which is rather large in the strong interaction scale, but are rather smaller than the experimental values.
The formalism was further applied to other hadron-nucleon systems.~{\cite{rf:KMN}\cite{rf:Koike}}

The following point, however, remained unclarified in those works.
In the analysis of the QCD sum rule for the hadron in the vacuum, the correlation function of the hadron interpolating field, whose matrix element is taken with respect to the vacuum, is considered.
The imaginary part of the correlation function consists of a pole term corresponding to the ground state and  a continuum term corresponding to the excited states.
Under the Borel transformed dispersion integral, the continuum contribution is exponentially suppressed compared to the ground state contribution due to the energy difference between the ground state and the continuum threshold.
For this reason the sum rule analysis is expected to be insensitive to the detailed form of the continuum, so that the continuum is usually parametrized in a very simple form.
When one deals with the hadron correlation function, whose matrix element is taken with respect to the one-nucleon state, the situation is different.
The energy of the continuum threshold is not higher than the pole energy.
Therefore, it is not clear if the Borel transformed dispersion integral is really dominated by the pole term or not.

A related question is the following.
It is known that there is a loosely bound state in the spin-triplet nucleon-nucleon channel and an almost bound state in the spin-singlet nucleon-nucleon channel.
If there is a zero-energy bound state, the scattering length diverges.
Therefore, the $NN$ scattering lengths are expected to be very sensitive to the $NN$ interaction strength.
On the other hand, it is hard to believe that the nucleon matrix elements of the quark-gluon operators would be very sensitive to the $NN$ interaction strength.
It seems strange that the sum rule relates these two quantities of very different natures.

Another point is that only the spin-averaged sum rules are obtained from the spin-averaged correlation function.
The $NN$ channel is special in the sense that selecting the isospin channel automatically selects the spin state.
In Ref.~\cite{rf:KM}, the sum rules for the spin-triplet and singlet scattering lengths are obtained by combining isospin states.
As far as the pole term is concerned, the above selection rule is correct,
but it does not hold for the continuum.
The sum rule in Ref.~3) is valid only if the pole term is dominant.
Therefore, it is more desirable to construct the spin-dependent sum rules.

In this paper we consider the spin-dependent correlation function of the nucleon interpolating field, where the matrix element is taken with respect to the spin-nonaveraged one-nucleon state.
The purpose of this paper is two-fold.
First, we extend the procedure of the sum rule to the case of the spin-nonaveraged correlation function.
Second, we show that the dispersion integral of the correlation function around the nucleon threshold can be regarded as a measure of the nucleon-nucleon interaction strength.
As a result, we derive sum rules which relate the spin-dependent $NN$ interaction strengths with the spin-dependent nucleon matrix elements of the quark-gluon operators.

\section{Formulation}
\subsection{Physical content of the correlation function and Borel sum rules}
Consider the spin-dependent correlation function, $\Pi(q \hat p s)$,
\begin{eqnarray*}
\Pi(q \hat p s)=-i\int d^4x e^{iqx}\langle\hat p s|T(\psi(x)\bar\psi(0))|\hat p s\rangle ,
\end{eqnarray*}
where  $|\hat p s\rangle $ is the one-nucleon state with momentum $\hat p$ and spin $s$ ($\hat p^2=\M^2$, $s^2=-1$ and $\hat ps=0$, where $\M$ is the nucleon mass) normalized as  $\langle\hat p s|\hat p' s' \rangle =(2\pi)^3\delta^3(\vec p-\vec p')\delta_{ss'}$ and $\psi$ is the normalized nucleon field operator $\langle 0|\psi(0)|\hat p s\rangle=u(ps)$, where $u(ps)$ is a positive energy solution of the free Dirac equation for the nucleon.
In this paper, momentum with $\;\hat{ }\;$ represents the nucleon on-shell momentum.
Later, the normalized nucleon field, $\psi$, is replaced by the unnormalized nucleon interpolating field (quark-gluon composite field), $\eta$.
The following discussion, however, holds as it is for the interpolating field, except for the normalization.
Naively, the dispersion relation for the correlation function, $\Pi(q \hat p s)$, is written as
\begin{eqnarray}\label{eq:dr}
\Pi(q \hat p s)=-{1 \over \pi}\int^\infty_{-\infty}dq'_0{1 \over q_0-q'_0+i\eta}{\rm Im}\Pi(q' \hat p s) ,
\end{eqnarray}
where $q'=(q'_0, \vec q)$.
Throughout this paper, whenever we take the imaginary part of a quantity, we approach the real energy axis from above in the complex energy plane.
Therefore, strictly speaking, ${\rm Im}\Pi$ is the imaginary part of the retarded correlation function.
The QCD sum rules are obtained by evaluating the left-hand side of Eq.~(\ref{eq:dr}) by the operator product expansion (OPE) and expressing the right-hand side in terms of physical quantities.

Let us consider the singularities of $\Pi(q \hat p s)$ as functions of $q_0$.
In the complex $q_0$ plane, $\Pi(q \hat p s)$ has a branch cut from the lowest $B=2$ continuum threshold to the right and another branch cut starting from the lowest $B=0$ continuum threshold to the left.
In addition, $\Pi(q \hat p s)$ has second-order poles at $q_0=\pm\sqrt{\vec q^2+M^2}\equiv\pm \E$ whose coefficients are the $NN$ and $N\bar N$ T-matrices $T_{+}$ and $T_{-}$, respectively:
\begin{eqnarray*}
&&T_{+}(\hat q r\hat p s;\hat q r\hat p s)\cr
&=&-i\int d^4x e^{iqx}\sqrt{M \over\E}\bar u(q r)(\gsl{q}-M)\langle\hat p s|T(\psi(x)\bar\psi(0))|\hat p s\rangle(\gsl{q}-M)\sqrt{M \over\E}u(q r)\cr
&=&(q_0-\E)^2{M \over\E}\bar u(q r)\Pi(q \hat p s)u(q r),\cr
&&T_{-}(\hat q r\hat p s;\hat q r\hat p s)\cr
&=&-i\int d^4x e^{iqx}\sqrt{M \over\E}\bar v(\bar q \bar r)(\gsl{q}-M)\langle\hat p s|T(\psi(x)\bar\psi(0))|\hat p s\rangle(\gsl{q}-M)\sqrt{M \over\E}v(\bar q\bar r)\cr
&=&(q_0+\E)^2{M \over\E}\bar v(\bar q \bar r)\Pi(q \hat p s)v(\bar q \bar r).
\end{eqnarray*}
Here $\bar q=(q_0,-\vec q)$, $\bar r=(r_0,-\vec r)$ and $v(\bar q \bar r)$ is the negative energy solution of the free Dirac equation for the nucleon.

In order to take out the pole contribution from ${\rm Im}\Pi(q \hat p s)$ it is convenient to define off-shell $NN$ and $N\bar N$ T-matrices by
\begin{eqnarray}\label{eq:offT}
&&T_{+}(q' r'\hat p' s';q r\hat p s)\cr
&=&-i\int d^4x e^{iq'x}\sqrt{M \over\EP}\bar u(q' r')(\gsl{q'}-M)\langle\hat p' s'|T(\psi(x)\bar\psi(0))|\hat p s\rangle(\gsl{q}-M)\sqrt{M \over\E}u(q r),\cr
&&T_{-}(q' r'\hat p' s';q r\hat p s)\cr
&=&-i\int d^4x e^{iq'x}\sqrt{M \over\E}\bar v(\bar q \bar r)(\gsl{q}-M)\langle\hat p' s'|T(\psi(x)\bar\psi(0))|\hat p s\rangle(\gsl{q'}-M)\sqrt{M \over\EP}v(\bar q' \bar r').\cr &&
\end{eqnarray}
Note that Eq.~(\ref{eq:offT}) is just a definition of the T-matrix off the mass shell, but the LSZ reduction formula shows rigorously that it is the T-matrix on the mass shell.

In order to separate the contribution from the poles at $q_0=\E$ and $q_0=-\E$, we introduce the projection operators $\Lambda_+$ and $\Lambda_-$ by
\begin{eqnarray*}
&&\Lambda_+(q s)=u(q s)\bar u(q s)={\gsl{\hat q}+M \over 2M}{1+\gamma_5\gsl{s} \over 2},\cr
&&\Lambda_-(q s)=v(q s)\bar v(q s)={\gsl{\hat q}-M \over 2M}{1+\gamma_5\gsl{s} \over 2},
\end{eqnarray*}
which have the properties
\begin{eqnarray*}
&&\Lambda_+^2(q s)=\Lambda_+(q s),\cr
&&\Lambda_-^2(q s)=\Lambda_-(q s),\cr
&&\Lambda_+(q s)\Lambda_+(q \bar s)=\Lambda_-(q s)\Lambda_-(q \bar s)=\Lambda_+(q s)\Lambda_-(q s')=0.
\end{eqnarray*}
Then we define the projected correlation functions by
\begin{eqnarray*}
\Pi_+(q r \hat p s)
&=&{\M\over\E}\tr\left\{\Lambda_+(\bar q \bar r)\Pi(q \hat p s)\right\},\cr
\Pi_-(q r \hat p s)
&=&{\M\over\E}\tr\left\{\Lambda_-(q r)\Pi(q \hat p s)\right\}.
\end{eqnarray*}

The projected correlation functions are related to the off-shell T-matrices as
\begin{eqnarray*}
\Pi_\pm(q r \hat p s)
={T_\pm(q r \hat p s) \over (q_0\mp\E)^2} ,
\end{eqnarray*}
where
$T_\pm(q r \hat p s) \equiv T_\pm(q r \hat p s;q r \hat p s)$.
Clearly, $\Pi_\pm(q r \hat p s)$ has a second-order pole at $q_0=\pm\E$ but not at $q_0=\mp\E$.

Naively, the dispersion relation for the projected correlation function, $\Pi_+$, is given by
\begin{eqnarray}\label{eq:PiI}
\Pi_+(q r \hat p s)
=-{1 \over \pi}\int^\infty_{-\infty}dq'_0{1 \over q_0-q'_0+i\eta}{\rm Im}\Pi_+(q' r \hat p s) .
\end{eqnarray}
Formally, the imaginary part of the correlation function is written as
\begin{eqnarray*}
{\rm Im}\Pi_+(q r \hat p s)&=&
{\rm Im}{1 \over \left(q_0 - \E + i\eta \right)^2}{\rm Re}T_+(q r \hat p s)+
{\rm Re}{1 \over \left(q_0 - \E + i\eta \right)^2}{\rm Im}T_+(q r \hat p s)\cr
&=& \pi\delta'(q_0 - \E) {\rm Re}T_+(q r \hat p s)+
{{\rm Pf} \over \left(q_0 - \E \right)^2}{\rm Im}T_+(q r \hat p s)\cr
&=& \pi\delta'(q_0 - \E) t - \pi\delta(q_0 - \E) u+
{{\rm Pf} \over \left(q_0 - \E \right)^2}{\rm Im}T_+(q r \hat p s)
,
\end{eqnarray*}
where
\begin{eqnarray*}
&&t = \left.{\rm Re}T_+(q r \hat p s)\right|_{q_0=\E} ,\cr
&&u = \left.{\partial \over \partial q_0}{\rm Re}T_+(q r \hat p s)\right|_{q_0=\E} .
\end{eqnarray*}
However, as it will turn out, when $\vec q = 0$ and $\vec p = 0$, the integral of the second term is divergent because $u$ is divergent, and the integral of the third term is also divergent because it behaves as $(q_0-\M)^{-{3 \over 2}}$ in the vicinity of $q_0 = \M$.
Therefore, Eq.~(\ref{eq:PiI}) is ill-defined.

Instead of Eq.~(\ref{eq:PiI}), we consider the dispersion relation for ${\gsl{q}-\M \over q_0}\Pi(qr\hat ps)$,
\begin{eqnarray*}
{\gsl{q}-\M \over q_0}\Pi(qr \hat ps)=-{1 \over \pi}\int^\infty_{-\infty}dq'_0{1 \over q_0-q_0'+i\eta}{\rm Im}\left\{{\gsl{q'}-\M \over q'_0}\Pi(q'r \hat ps)\right\} ,
\end{eqnarray*}
or for ${q_0-\E \over q_0}\Pi_+(qr \hat ps)$ in terms of the projected correlation function,
\begin{eqnarray}\label{eq:PiII}
&&{q_0-\E \over q_0}\Pi_+(qr \hat ps)=-{1 \over \pi}\int^\infty_{-\infty}dq'_0{1 \over q_0-q_0'+i\eta}{\rm Im}\left\{{q'_0-\E \over q'_0}\Pi_+(q'r \hat ps)\right\}.\qquad
\end{eqnarray}
Symmetrizing Eq.~(\ref{eq:PiII}) we obtain
\begin{eqnarray}\label{eq:QSR}
 {q_0-\E \over 2q_0}\Pi_+(qr \hat ps)
 +(q_0\rightarrow-q_0)
={1 \over \pi}\int^\infty_{-\infty}dq'_0{1 \over (q_0+i\eta)^2-q'^2_0}
(q'_0-\E){\rm Im}\Pi_+(q'r \hat ps),\cr
\end{eqnarray}
where
\begin{eqnarray}\label{eq:ImPi}
(q_0-\E){\rm Im}\Pi_+(q r \hat p s)
= -\pi\delta(q_0 - \E) t +
{{\rm P} \over q_0 - \E}{\rm Im}T_+(q r \hat p s)
.
\end{eqnarray}
Now Eq.~(\ref{eq:QSR}) is well-defined when $\vec q = 0$ and $\vec p = 0$ because $u$ does not appear and the second term behaves as $(q_0-\M)^{-{1 \over 2}}$ in the vicinity of $q_0 = \M$.
Applying the Borel transformation,
\begin{eqnarray*}
       L_B\equiv
       \lim_{{n\rightarrow\infty \atop -q_0^2\rightarrow\infty}
       \atop -q_0^2/n = \MBS}
       {(q_0^2)^n\over(n-1)!}\left(-{d\over dq_0^2}\right)^n ,
\end{eqnarray*}
to both sides of Eq.~(\ref{eq:QSR}), we obtain
\begin{eqnarray}\label{eq:BSR}
& &L_B\Big[{q_0-\E \over 2q_0}\Pi_+(qr \hat ps)
          +(q_0\rightarrow-q_0)
      \Big]\cr
&=&-{1 \over \pi}\int^\infty_{-\infty}dq'_0
    {1\over\MBS}\exp\left(-{q_0'^2\over\MBS}\right)
    (q'_0-\E){\rm Im}\Pi_+(q' r \hat p s) ,
\end{eqnarray}
where $\MB$ is the Borel mass.
In order to derive the Borel sum rules we must evaluate the left-hand side by the OPE and parametrize the right-hand side in terms of physical quantities.
Now, the question is how to parametrize the integrand of the right-hand side of Eq.~(\ref{eq:BSR}).

Let us recall the QCD sum rule for the nucleon in the vacuum, where the imaginary part of the correlation function has the form
\begin{eqnarray*}
{\rm Im}\Pi_{+}(q)
\propto -\pi\delta(q_0 -\E)+\sigma(q) .
\end{eqnarray*}
The first term is the contribution from the nucleon pole term, and the second term is due to the excited states.
Borel transformation on the dispersion integral of ${\rm Im}\Pi_{+}(q)$ gives
\begin{eqnarray*}
& &L_B\Big[{1\over2q_0}\Pi_{+}(q)+(q_0\rightarrow-q_0)\Big]\cr
&=&-{1 \over \pi}\int dq_0'{1\over\MBS}e^{-q_0'^2/\MBS}{\rm Im}\Pi_{+}(q')\cr
&\propto&{1\over\MBS}e^{-\E^2/\MBS}
   -{1\over\pi}\int^\infty_{\omega} dq_0'{1\over\MBS}e^{-q_0'^2/\MBS}\sigma'(q') ,
\end{eqnarray*}
where the second term, the contribution from the excited states, starts at the continuum threshold, $\omega$ ($\omega > \E$), and is exponentially suppressed compared to the first term.
For this reason, it is possible to use the rough model of the hadron continuum,
\begin{eqnarray}\label{eq:piv}
{\rm Im}\Pi_{+}(q)
=-\lambda^2\pi\delta(q_0 -\E)
+\left\{\theta(q_0-\w)+\theta(-\w-q_0)\right\}{\rm Im}\Pi^{OPE}_{+}(q) ,\qquad
\end{eqnarray}
where $\Pi^{OPE}_{+}$ is the asymptotic form of the correlation function in the OPE, $\w$ is the effective continuum threshold, and the normalization constant $\lambda$ is explicitly included ($\langle 0|\eta(0)|\hat p s\rangle=\lambda u(ps)$).

Let us turn to the problem at hand. 
As an extension of Eq.~(\ref{eq:ImPi}) one might parametrize $(q_0-\E){\rm Im}\Pi_+$ as
\begin{eqnarray}\label{eq:pin}
&&(q_0-\E){\rm Im}\Pi_+\cr
&=&-\lambda^2\pi\delta(q_0-\E)t
+\left\{\theta(-q_0-\omega_-)+\theta(q_0-\omega_+)\right\}(q_0-\E){\rm Im}\Pi_+^{OPE},\qquad
\end{eqnarray}
by approximating the second term of the right-hand side of Eq.~(\ref{eq:pin}) by its asymptotic form.
However, the second term starts at $q_0=\omega=\sqrt{4\M^2+{\vec q}^2}-\M$ ($\omega \leq \sqrt{\M^2+\vec q^2}$), and it is not exponentially suppressed compared to the first term.
Therefore, one cannot justify Eq.~(\ref{eq:pin}).
One has to know the behavior of the second term around the threshold.
For this purpose it is important to note that the off-shell optical theorem holds for $T$.
When the center-of-mass energy is above the threshold of the $NN$ channel and below the threshold of the next channel, only the $NN$ states contribute in the intermediate states, and the off-shell optical theorem is simplified as
\begin{eqnarray}\label{eq:OT}
{\rm Im}T_+(q \hat p;q \hat p)
=-\pi\int {d^3p_n\over(2\pi)^3}{d^3q_n\over(2\pi)^3}(2\pi)^3
\delta^4(\hat p+q-\hat p_n-\hat q_n)
T_+(q \hat p;\hat q_n\hat p_n)
T_+(\hat q_n\hat p_n;q \hat p).\cr
\end{eqnarray}
In order to simplify the notation we introduce the scattering amplitude $f$ by
\begin{eqnarray*}
f(q'p';qp) = -{\mu'^{1/2}\mu^{1/2} \over 2\pi}T_+(q'p';qp) ,
\end{eqnarray*}
where $\mu={q_0p_0\over q_0+p_0}$ and $\mu'={q'_0p'_0\over q'_0+p'_0}$.
Moreover, we  go to the center-of-mass frame ($\vec q + \vec p = \vec q' + \vec p' =0$) and restrict ourselves to the $s$-wave.
We define three scattering amplitudes, $f_0$, $f_1$ and $f_2$ as
\begin{eqnarray*}
f_0(k) = f(\hat q'\hat p';\hat q\hat p) ,
\end{eqnarray*}
where $|\vec p|=|\vec q|=|\vec p'|=|\vec q'|=k$, $p_0=q_0=p'_0=q'_0=\sqrt{\M^2+k^2}$,
\begin{eqnarray*}
f_1(k) = f(q'\hat p';\hat q\hat p) ,
\end{eqnarray*}
where $|\vec p|=|\vec q|=k$, $|\vec p'|=|\vec q'|=0$, $p_0=q_0=\sqrt{\M^2+k^2}$, $p'_0=\M$, $q'_0=2\sqrt{\M^2+k^2}-\M$, and
\begin{eqnarray*}
f_2(k) = f(q'\hat p';q \hat p) ,
\end{eqnarray*}
where $|\vec p|=|\vec q|=|\vec p'|=|\vec q'|=0$,
$p_0=p'_0=\M$, $q_0=q'_0=2\sqrt{\M^2+k^2}-\M$.

It is well known that the on-shell scattering amplitude $f_0$ has the form
\begin{eqnarray*}
f_0(k) = {1 \over -ik + k\cot\delta} = {1 \over -ik + {1 \over a}+{1 \over 2}rk^2+O(k^4)},
\end{eqnarray*}
where $a$ is the scattering length and $r$ the effective range.
Similarly, the off-shell scattering amplitude $f_2$ has the form
\begin{eqnarray}\label{eq:fii}
f_2(k) = {1 \over i\left\{-k+bk^3+O(k^5)\right\}+
\left\{{1 \over a} + {1 \over 2}\tilde r k^2 + O(k^4)\right\}},
\end{eqnarray}
which can be shown as follows.
First, since the discontinuity of the T-matrix along the real energy axis is proportional to the imaginary part of the T-matrix, the Taylor expansion of the real and imaginary parts of the scattering amplitude includes only even and odd powers of $k$, respectively.
Second, $f_0$, $f_1$ and $f_2$ coincide on the mass-shell ($k=0$),
$f_0(0)=f_1(0)=f_2(0)=a$.
Therefore, we have
\begin{eqnarray}\label{eq:Ref}
&&{\rm Re}{1 \over f_0(k)}= {1 \over a} + {1 \over 2} r k^2 + O(k^4) ,\cr
&&{\rm Re}{1 \over f_2(k)}= {1 \over a} + {1 \over 2} \tilde r k^2 + O(k^4) .
\end{eqnarray}
Third, from Eq.~(\ref{eq:OT}) the following relations hold,
\begin{eqnarray*}
&&{\rm Im}f_0(k)=k|f_0(k)|^2 ,\cr
&&{\rm Im}f_2(k)=k|f_1(k)|^2 .
\end{eqnarray*}
Therefore, we have
\begin{eqnarray}\label{eq:Imf}
&&{\rm Im}{1 \over f_0(k)}=-{{\rm Im}f_0(k) \over |f_0(k)|^2} = -k ,\cr
&&{\rm Im}{1 \over f_2(k)}=-{{\rm Im}f_2(k) \over |f_2(k)|^2} = -k{|f_1(k)|^2 \over |f_2(k)|^2} = -k + bk^3 + O(k^5) .
\end{eqnarray}
Equation~(\ref{eq:fii}) follows from Eqs.~(\ref{eq:Ref}) and (\ref{eq:Imf}).

It should be noted that $\tilde r$ is different from the effective range $r$,
but $\tilde r$ coincides with $r$ in the limit $a \rightarrow \infty$:
\begin{eqnarray*}
\tilde r = r + O\left({1 \over a}\right).
\end{eqnarray*}
This is shown as follows.
Equations~(\ref{eq:Ref}) and (\ref{eq:Imf}) indicate that both $O(1)$ and $O(k)$ terms of ${1/f_0(k)}$ and ${1/f_2(k)}$, which are real and imaginary respectively, coincide with each other.
Therefore, we have
\begin{eqnarray}\label{eq:ratio}
{f_2(k) \over f_0(k)}=1+O(k^2) .
\end{eqnarray}
Equation~(\ref{eq:ratio}) is independent of $a$ and therefore holds also in the limit $a \rightarrow \infty$ due to the continuity.
Since ${1/f_0(k)}$ and ${1/f_2(k)}$ do not have the $O(1)$ terms in this limit, they must coincide with each other up to the $O(k^2)$ terms, i.e. $\tilde r = r + O\left({1 \over a}\right)$.

From Eq.~(\ref{eq:fii}) we have
\begin{eqnarray*}
{\rm Re}f_2(k)&=&\left\{ 
        \begin{array}{ll}a + a^2\kappa + O(\kappa^2) &q_0 < \M \\
        a-\left(1+{\tilde r \over 2a}\right)a^3k^2 + O(k^4) &q_0 > \M
        \end{array} \right. ,\\
{\rm Im}f_2(k)&=&\left\{
        \begin{array}{ll}0 &q_0 < \M \\
        a^2 k + O(k^3) &q_0 > \M
        \end{array} \right. .
\end{eqnarray*}
where $\kappa=-ik$. One sees that
\begin{eqnarray*}
\left.{\partial \over \partial q_0}{\rm Re}T_+\right|_{q_0=\M} = -{1 \over 2\pi}\left.{\partial \over \partial q_0}\left\{{q_0+\M \over q_0\M}{\rm Re}f_2\right\}\right|_{q_0=\M}=\infty,
\end{eqnarray*}
and
\begin{eqnarray*}
{\rm Im}T_+=-{1 \over 2\pi}{q_0+\M \over q_0\M}{\rm Im}f_2\propto (q_0-M)^{1 \over 2},
\end{eqnarray*}
which make the naive dispersion relation, Eq.~(\ref{eq:PiI}), ill-defined.

Having understood the structure of ${\rm Im}T_+$, we proceed to the integral of the right-hand side of Eq.~(\ref{eq:BSR}), $I$, in the vicinity of $q_0 = \M$,
\begin{eqnarray*}
I&=&-{1 \over \pi}\int_{\vic\M}dq'_0
    {1\over\MBS}\exp\left(-{q_0'^2\over\MBS}\right)
    (q'_0-\M){\rm Im}\Pi_+.
\end{eqnarray*}
The integral, $I$, can be decomposed as
\begin{eqnarray}\label{eq:III}
 I=I_{t}+I_{c}\ (+I_{b}).
\end{eqnarray}
In Eq.~(\ref{eq:III}), the first term, $I_{t}$, is the threshold contribution, given by
\begin{eqnarray*}
I_{t}=-{1\over\MBS}\exp\left(-{\M^2\over\MBS}\right){4\pi a \over \M}.
\end{eqnarray*}
The second term, $I_{c}$, is the continuum contribution, given by
\begin{eqnarray*}
I_{c}=-{1 \over \pi}\int_{\vic\M} dq'_0{1\over\MBS}\exp\left(-{q_0'^2\over\MBS}\right)
{{\rm P} \over q'_0-\M}\left\{-{2\pi}{q'_0+\M\over q'_0\M}{\rm Im}f^{cut}_2\right\},
\end{eqnarray*}
where
\begin{eqnarray*}
{\rm Im}f^{cut}_2 = {k\over {1 \over a^2} + \left(1+{\tilde r\over a}+{b\over a^2}\right)k^2+O(k^4)}\theta(q_0-M).
\end{eqnarray*}
The last term, $I_{b}$, is the bound-state contribution, which has to be taken into account if there is a bound state, given by
\begin{eqnarray*}
I_{b}=-{1 \over \pi}\int_{\vic\M} dq'_0{1\over\MBS}\exp\left(-{q_0'^2\over\MBS}\right)
{{\rm P} \over q'_0-\M}\left\{-{2\pi}{q'_0+\M\over q'_0\M}{\rm Im}f^{pole}_2\right\},
\end{eqnarray*}
where
\begin{eqnarray*}
{\rm Im}f^{pole}_2 
 &=& -i\pi\left\{\left.{\partial\over\partial k}\left({1 \over f_2}\right)\right|_{k=i\kappa_0}\right\}^{-1}\delta(\kappa-\kappa_0)\cr
 &\equiv& \pi c \delta(\kappa-\kappa_0) ,
\end{eqnarray*}
and $i\kappa_0$ is the pole momentum, $1/f_2(i\kappa_0)=0$.

By performing the integral, the continuum contribution becomes
\begin{eqnarray*}
I_c
&\approx&
-{1 \over \pi}\int dq'_0
{1\over\MBS}\exp\left(-{q_0'^2\over\MBS}\right)
\left\{{{\rm P} \over q'_0-\M}\left(-2\pi{q'_0+\M \over q'_0\M}\right)
{k \over {1 \over a^2}+\left(1+{\tilde r \over a}+{b\over a^2}\right)k^2}\right\}\cr
&=&
{1\over\MBS}\exp\left(-{\M^2\over\MBS}\right)
{4\pi|a|\over\sqrt{{1\over a^2}+\M^2\left(1+{\tilde r \over a}+{b\over a^2}\right)}},
\end{eqnarray*}
which is simplified in two limits of $a$ as
\begin{eqnarray*}
I_c\rightarrow\left\{
\begin{array}{ll}
{1\over\MBS}\exp\left(-{\M^2\over\MBS}\right)
   {4\pi\over\sqrt{1+\M^2 b}} a^2, &(a \rightarrow 0)\cr
{1\over\MBS}\exp\left(-{\M^2\over\MBS}\right)
   {4\pi \over \M}|a|\left(1-{r\over 2a}\right)+O\left({1 \over a}\right).
   &(a \rightarrow \infty) 
\end{array}
\right.
\end{eqnarray*}

Similarly, the bound-state contribution becomes
\begin{eqnarray*}
I_b&=&-{1 \over \pi}\int_{\sim\M} dq'_0{1\over\MBS}\exp\left(-{q_0'^2\over\MBS}\right)\left\{{{\rm P} \over q'_0-\M}\left(-2\pi{q'_0+\M \over q'_0\M}\right)
\pi c\delta(\kappa-\kappa_0)
\right\}\cr
&=&-{1 \over \pi}{2\kappa_0\over\sqrt{M^2-\kappa_0^2}}
{1\over\MBS}\exp\left(-{\omega'^2\over\MBS}\right)
{1 \over \omega' - \M}2\pi^2{M+\omega'\over M\omega'}
\pi c,
\end{eqnarray*}
where $\omega'=2\sqrt{\M^2-\kappa_0^2}-\M$.
In the limit $a\rightarrow\infty$,
\begin{eqnarray*}
&&\kappa_0 \rightarrow -{1 \over a}+O\left({1 \over a^2}\right),\cr
&&c \rightarrow \left(1-{r \over a}\right)+O\left({1 \over a^2}\right),
\end{eqnarray*}
and the bound-state contribution is simplified as
\begin{eqnarray*}
I_b\rightarrow
{1\over\MBS}\exp\left(-{\M^2\over\MBS}\right){8\pi \over \M}a\left(1-{r \over 2a}\right)+O\left({1 \over a}\right).
\qquad (a \rightarrow \infty)
\end{eqnarray*}

Let us suppose that one can freely change the interaction strength of two nucleons and examine how the integral $I$ should change as a function the interaction strength.
When the interaction is weak, the scattering length is also small, and the integral $I$ is dominated by $I_{t}$:
\begin{eqnarray*}
I&=&I_{t}+I_{c}\cr
 &=&-{1 \over \MBS}\exp\left(-{\M^2\over\MBS}\right){4\pi \over \M}a+O(a^2) .
\end{eqnarray*}
As the interaction becomes stronger, the scattering length increases and the integral, $I$, also increases.
As the interaction strength increases further, the scattering length eventually diverges when the bound state is just formed.
Just before the bound state is formed, the integral $I$ becomes
\begin{eqnarray*}
I&=&I_{t}+I_{c}\cr
&=&-{1 \over \MBS}\exp\left(-{\M^2\over\MBS}\right)\left\{{4\pi a \over \M}-
{4\pi a \over \M}\left(1-{r \over 2a}\right)+O\left({1 \over a}\right)\right\}\cr
&=&-{1 \over \MBS}\exp\left(-{\M^2\over\MBS}\right){2\pi r \over \M}+O\left({1 \over a}\right),
\end{eqnarray*}
and just after the bound state is formed, it becomes
\begin{eqnarray*}
I&=&I_{t}+I_{c}+I_{b}\cr
&=&-{1 \over \MBS}\exp\left(-{\M^2\over\MBS}\right)\left\{{4\pi a \over \M}-
{-4\pi a \over \M}\left(1-{r \over 2a}\right)+
{-8\pi a \over \M}\left(1-{r \over 2a}\right)+O\left({1 \over a}\right)\right\}\cr
&=&-{1 \over \MBS}\exp\left(-{\M^2\over\MBS}\right){2\pi r \over \M}+O\left({1 \over a}\right) .
\end{eqnarray*}
This shows that before and after the bound state is formed the integral is continuous, though the scattering length diverges with opposite signs.
This observation leads us to conjecture that the integral around the threshold is a measure of the $NN$ interaction strength.
Based on this conjecture we define the $NN$ interaction strength, $\alpha$, by
\begin{eqnarray}\label{eq:I}
I&=&-{1 \over \pi}\int_{\vic\M}dq'_0
    {1\over\MBS}\exp\left(-{q_0'^2\over\MBS}\right)
    (q'_0-\M){\rm Im}\Pi_+\cr
&\equiv&-{1\over\MBS}\exp\left(-{\M^2\over\MBS}\right){4\pi \alpha \over \M}.
\end{eqnarray}

In the dispersion integral, the imaginary part of the correlation function,  ${\rm Im}\Pi_+$, contains the contribution from all possible intermediate states such as those of the $NN$, $NN\pi$ channels and so on.
However, only the $NN$ channel contributes around the threshold.
We assume that the contribution from the $NN$ state is taken into account by the form of the right-hand side of Eq.~(\ref{eq:I}) and that the rest is approximated by the asymptotic form of the correlation function starting from an (effective) threshold, $\omega_+$,  for the $B=2$ channels other than the $NN$ channel and $\omega_-$ for the $B=0$ channels:
\begin{eqnarray}\label{eq:ImPi}
& &(q_0-\M){\rm Im}\Pi_+ \cr
&=&\lambda^2\pi\delta(q_0-\M){2\pi\over\mu}\alpha 
 +\left\{\theta(-q_0-\omega_-)+\theta(q_0-\omega_+)\right\}(q_0-\M){\rm Im}\Pi_+^{OPE},\qquad\quad
\end{eqnarray}
where the normalization constant $\lambda$ is explicitly included. 
This is possible now because the contribution from states other than those of the $NN$ channel is exponentially suppressed compared to the $NN$ contribution. 

\subsection{OPE of the correlation function and results}
Let us turn to the OPE.
We take the interpolating field of the neutron as~\cite{rf:Ioffe}
\begin{eqnarray*}
  \eta(x) =  \epsilon_{abc}
  \left (d^{Ta}(x) C\gamma_\mu d^b(x)\right )
  \gamma_5\gamma^\mu u^c(x) ,
\end{eqnarray*}
where $C$ denotes the charge conjugation operator and $a$, $b$ and $c$ are color indices.
We take into account all the operators of dimension less than or equal to four.
We also include four-quark operators, they are of  dimension six.
The operators which involve the quark mass are ignored.
In the OPE, the neutron correlation function, where the matrix elements are taken with respect to the one-nucleon state, is given as
\begin{eqnarray}\label{eq:OPE}
  & & \Pi^{OPE}(q\hat ps)\cr\cr
  &=&{1\over4\pi^4}\gamma^\mu\Big[ 
    q^2\ln(-q^2)\pi^2\Big\{-{7\over3}\langle\bar d\gamma_\mu d\rangle_N
                           -{1\over3}\langle\bar u\gamma_\mu u\rangle_N\Big\}\cr
  & &\qquad
   +q_\mu q^\nu\ln(-q^2)\pi^2\Big\{-{2\over3}\langle\bar d\gamma_\nu d\rangle_N
                           -{2\over3}\langle\bar u\gamma_\nu u\rangle_N\Big\}\cr
  & &\qquad
   +q_\mu\ln(-q^2)\pi^2\Big\{
    {1\over8}\left\langle{\alpha_s\over\pi}G^{\alpha\beta}G_{\alpha\beta}
             \right\rangle_N\Big\}\cr
  & &\qquad
   +q^\nu\ln(-q^2)\pi^2\Big\{
    -{1\over6}\left\langle{\alpha_s\over\pi}{\cal S}
                  [G^{\rho}_{\mu}G_{\rho\nu}]\right\rangle_N \cr
  & &\qquad  +{16\over3}i\langle{\cal S}[\bar d\gamma_\mu D_\nu d]\rangle_N
    +{4\over3}i\langle{\cal S}[\bar u\gamma_\mu D_\nu u]\rangle_N\Big\}\cr
  & &\qquad
   +q_\mu{1\over q^2}\pi^4
    \Big\{{8\over3}\langle\bar d d 
                          \bar d d\rangle_N \Big\} \Big] \cr
  &+&{1\over4\pi^4}\Big[ q^2\ln(-q^2)\pi^2\{-\langle\bar uu\rangle_N\}
     +q^\mu{1\over q^2}\pi^4\Big\{{16\over3}
      \langle\bar uu 
             \bar d\gamma_\mu d\rangle_N\Big\} 
 \Big] \cr
  &+&{1\over4\pi^4}\gamma^\mu\gamma_5\Big[ 
    q^2\ln(-q^2)\pi^2\Big\{{5\over3}\langle\bar d\gamma_\mu\gamma_5 d\rangle_N
                  -{1\over3}\langle\bar u\gamma_\mu\gamma_5 u\rangle_N\Big\}\cr
  & &\qquad
   +q_\mu q^\nu\ln(-q^2)\pi^2\Big\{
          -{2\over3}\langle\bar d\gamma_\nu\gamma_5 d\rangle_N
          -{2\over3}\langle\bar u\gamma_\nu\gamma_5 u\rangle_N\Big\}\cr
  & &\qquad
   +q^\nu\ln(-q^2)\pi^2\Big\{
          -{8\over3}\langle{\cal S}[\bar d\gamma_\mu\gamma_5 iD_\nu d]\rangle_N
          +{4\over3}\langle{\cal S}[\bar u\gamma_\mu\gamma_5 iD_\nu u]\rangle_N\Big\}\cr
  & &\qquad
   +q^\nu{1\over q^2}\pi^4\Big\{
           -{16\over3}\langle\bar d d 
             \bar d\gamma_5i\sigma_{\mu\nu} d\rangle_N\Big\}
 \Big] \cr
  &+&{1\over4\pi^4}\gamma_5\sigma^{\mu\nu}\Big[
    q^2\ln(-q^2)\pi^2\Big\{
  -{1\over6}\langle\bar u\gamma_5\sigma_{\mu\nu}u\rangle_N \Big\}
   +q_\mu q^\rho\ln(-q^2)\pi^2\Big\{
  -{2\over3}\langle\bar u\gamma_5\sigma_{\nu\rho}u\rangle_N \Big\}
 \Big\}\cr
  & &\qquad
   +q_\mu{1\over q^2}\pi^4\Big\{{16\over3}i
    \langle\bar uu 
           \bar d\gamma_\nu\gamma_5 d\rangle_N \Big\} \Big] .
\end{eqnarray}
In Eq.~(\ref{eq:OPE}), $D$ is the covariant derivative, 
${\cal S}[A_\mu B_\nu]\equiv A_{\{\mu} B_{\nu\}}-({\rm traces})$
, where $_\{\quad_\}$ represents symmetrization over the Lorentz indices and $-({\rm traces})$ represents the subtraction of the trace terms.
Here, $\langle{\cal O}\rangle_N$ is the connected part of nucleon matrix element of ${\cal O}$,
$\langle{\cal O}\rangle_N\equiv\langle N|{\cal O}|N \rangle-\langle N|N \rangle\langle{\cal O}\rangle_0$, where $\langle{\cal O}\rangle_0$ represents the vacuum expectation value of ${\cal O}$, and $|N \rangle\equiv |\hat ps\rangle$.

Substituting the projection of Eq.~(\ref{eq:OPE}) into the left-hand side and Eq.~(\ref{eq:ImPi}) into the right-hand side of Eq.~(\ref{eq:BSR}), respectively,
and splitting $\alpha$ into spin-independent and spin-dependent parts, $\alpha=\alpha^{indep}+\alpha^{dep}$, we obtain sum rules for the spin independent and dependent interaction strengths, respectively, as
\begin{eqnarray}\label{eq:SRindep}
 & &-\lambda^2 {4\pi\over\M}\alpha^{indep}{1\over\MBS}\exp\left(-{\M^2\over\MBS}\right)\cr
 &=&{1\over4\pi^4}\Big\{
    \left(C_2\M\MB -C_3\MBS\right)
     \Big[\pi^2\Big\{-3\langle d^\dagger d\rangle_N
      -\langle u^\dagger u\rangle_N \Big\}
       -\pi^2\langle\bar uu\rangle_N \Big]\cr
 & &\qquad
    +\left(C_1\M-C_2\MB\right)
     \Big[\pi^2\Big\{
    {1\over8}\left\langle{\alpha_s\over\pi}G^{\alpha\beta}G_{\alpha\beta}
             \right\rangle_N
    -{1\over6}\left\langle{\alpha_s\over\pi}{\cal S} 
                  [G^{\rho}_{0}G_{\rho0}]\right\rangle_N\cr
 & &\qquad
    +{16\over3}i\langle{\cal S}[\bar d\gamma_0 D_0 d]\rangle_N
    +{4\over3}i\langle{\cal S}[\bar u\gamma_0 D_0 u]\rangle_N\Big\}\Big]\cr
 & &\qquad
    +{\M\over\MBS}\Big[{8\over3}\pi^4\langle\bar dd 
                                                    \bar dd\rangle_N
    +{16\over3}\pi^4\langle\bar uu 
                                   d^\dagger d\rangle_N\Big]
 \Big\} ,
\end{eqnarray}
and
\begin{eqnarray}\label{eq:SRdep}
 & &-\lambda^2 {4\pi\over\M}\alpha^{dep}{1\over\MBS}\exp\left(-{\M^2\over\MBS}\right)\cr
 &=&{1\over4\pi^4}r_k\Big\{
    \left(C_2\M\MB-C_3\MBS\right)\cr
 & &\qquad\times 
    \Big[\pi^2\Big\{{5\over3}\langle\bar d\gamma_k\gamma_5 d\rangle_N
               -{1\over3}\langle\bar u\gvec\gamma_5 u\rangle_N\Big\}
     -i{1\over3}\pi^2\langle\bar u\gamma_5\sigma_{k0} u\rangle_N
 \Big]\cr
 & &\qquad
     +\left(C_1\M-C_2\MB\right)\cr
 & &\qquad\times
    \Big[\pi^2\Big\{-{8\over3}\langle\bar d\gamma_k\gamma_5 iD_0 d\rangle_N
               +{4\over3}\langle\bar u\gamma_k\gamma_5 iD_0 u\rangle_N\Big\}
\Big]\cr
 & &\qquad
    +{\M\over\MBS}\Big[
    -{16\over3}\pi^4\langle\bar dd 
                                   \bar d\gamma_5i\sigma_{k0} d\rangle_N
    -{16\over3}\pi^4\langle\bar uu 
                                   \bar d\gamma_k\gamma_5 d\rangle_N
 \Big] \Big\} ,
\end{eqnarray}
where
\begin{eqnarray*}
  C_1
  &=&1-{1\over2}\left[\exp\left(-{\wp^2\over\MBS}\right)
                     +\exp\left(-{\wm^2\over\MBS}\right)\right],\cr
  C_2
  &=&-{1\over2}\left[{\wp\over\MB}\exp\left(-{\wp^2\over\MBS}\right)
               -{\wm\over\MB}\exp\left(-{\wm^2\over\MBS}\right)\right]
     +{\sqrt{\pi}\over4}\left[\Phi\left({\wp\over\MB}\right)
                             -\Phi\left({\wm\over\MB}\right)\right],\cr
  C_3
  &=&1-{1\over2}\left[
     \left(1+{\wp^2\over\MBS}\right)\exp\left(-{\wp^2\over\MBS}\right)
    +\left(1+{\wm^2\over\MBS}\right)\exp\left(-{\wm^2\over\MBS}\right)\right] ,
\end{eqnarray*}
and $\Phi$ is the error function.

From the QCD sum rule for the nucleon in the vacuum in Ref.~\cite{rf:Ioffe}, the normalization constant $\lambda^2$ is related with the vacuum expectation values of the operators according to
\begin{eqnarray*}
 & &\lambda^2 {1\over\MBS}\exp\left(-{\M^2\over\MBS}\right)\cr
 &=&{1\over4\pi^4}\Big[
  - D_2\MBQ{1\over8}
  - D_1\Big\{
    {\pi^2\over8}
    \left\langle{\alpha_s\over\pi}G^{\alpha\beta}G_{\alpha\beta}\right\rangle_0
       \Big\}
  - {1\over\MBS}\Big\{\pi^4{8\over3}\langle\bar dd\bar dd\rangle_0\Big\}\Big] ,
\end{eqnarray*}
where
\begin{eqnarray*}
  D_1&=&1-\exp\left(-{\w^2\over\MBS}\right),\cr
  D_2&=&1-\left(1+{\w^2\over\MBS}+{1\over2}{\w^4\over\MBQ}\right)
              \exp\left(-{\w^2\over\MBS}\right).
\end{eqnarray*}

The sum rule for the spin-independent part of the interaction strength, $\alpha^{indep}$, is nothing but the sum rule for the $NN$ scattering length in Ref.~\cite{rf:KM}, which was obtained starting from the spin-averaged correlation function. The difference, however, is that the scattering length in Ref.~\cite{rf:KM} is replaced by the interaction strength in Eq.~(\ref{eq:SRindep}).

Equation~(\ref{eq:SRdep}) is the new sum rule for the spin-dependent part of the interaction strength, $\alpha^{dep}$.
The new sum rule provides us with a relation between the spin-dependent $NN$ interaction strength and the nucleon matrix elements of the spin-dependent quark-gluon operators.
Therefore, if one knows the nucleon matrix elements of the spin-dependent quark-gluon operators, one can predict the spin-dependent $NN$ interaction strength, and vice versa.
In this paper we investigate the first possibility.

We now discuss the nucleon matrix elements of the quark-gluon operators.
Dimension-three operators are quark operators, $\bar q q$, $\bar q\gamma_\mu q$, $\bar q\gamma_\mu\gamma_5 q$ and $\bar q\gamma_5\sigma_{\mu\nu} q$.
The nucleon matrix elements, $\langle\bar q q\rangle_N$ and $\langle\bar q\gamma_\mu q\rangle_N$, are spin-independent and have already been discussed in Ref.~7),
while $\langle\bar q\gamma_\mu\gamma_5 q\rangle_N$ and $\langle\bar q\gamma_5\sigma_{\mu\nu} q\rangle_N$ are spin-dependent and are written in terms of the axial charge, $\Delta q$, and the tensor charge, $\delta q$, as
\begin{eqnarray*}
    &&\langle\bar q\gamma_\mu\gamma_5 q\rangle_N=\Delta q s_\mu ,\cr
    &&\langle\bar qi\gamma_5\sigma_{\mu\nu}q\rangle_N
    =\delta q(s_\mu \hat p_\nu-s_\nu \hat p_\mu)/\hat p_0 .
\end{eqnarray*}
The axial charge and the tensor charge are related to the structure functions~\cite{rf:JJ} as
\begin{eqnarray*}
  &&\Delta q=\int_0^1dx[g_1(x)+\bar g_1(x)],\cr
  &&\delta q=\int_0^1dx[h_1(x)-\bar h_1(x)],
\end{eqnarray*}
where $g_1$ is the longitudinal quark-spin distribution in the nucleon, $\bar g_1$ is the antiquark-spin distribution, $h_1$ is the quark-transversity distribution, and $\bar h_1$ is the antiquark-transversity distribution.
$h_1$, together with the unpolarized quark distributions $f_1$ and $g_1$, forms a complete set of twist-2 structure functions.

Recently $\Delta q$ has received much attention in connection with the spin content of the nucleon.
While a naive quark model picture suggests that the nucleon spin is carried by quarks, i.e. $\Delta u+\Delta d=1$, it was experimentally found that $\Delta u+\Delta d<<1$.~\cite{rf:EMC} 
In Ref.~10), $\Delta q$ for the proton was determined from recent EMC/SMC and SLAC data with the SU(3) symmetry and the hyperon $\beta$ decay as
\begin{eqnarray*}
   &&\Delta u = 0.83 \pm 0.03,\cr
   &&\Delta d = -0.43 \pm 0.03,
\end{eqnarray*}
at the renormalization scale $Q^2=10\ {\rm GeV}^2$.
There is also a lattice QCD calculation~\cite{rf:FKOU} with the result
\begin{eqnarray*}
   &&\Delta u = 0.638 \pm 0.054,\cr
   &&\Delta d = -0.347 \pm 0.046
\end{eqnarray*}
at $Q^2=2\ {\rm GeV}^2$.

On the other hand, up to now no experimental information is available for $\delta q$, since $h_1$ and $\bar h_1$ cannot be measured by deep inelastic scattering.
These quantities can, however, be measured by Drell-Yang processes, which are planned in a future RHIC experiment.
However, $\delta q$ for the proton was calculated on the lattice~\cite{rf:ADHK} with the result
\begin{eqnarray*}
  &&\delta u=0.839 \pm 0.060,\cr
  &&\delta d=-0.231 \pm 0.055
\end{eqnarray*}
at $Q^2=2\ {\rm GeV}^2$.
In this paper we use the lattice results both for $\Delta q$ and $\delta q$ and
ignore the $Q^2$ evolution of these matrix elements between $Q^2=1\ {\rm GeV}^2$ and $2\ {\rm GeV}^2$. 

Dimension-four operators are gluon operators, ${\alpha_s\over\pi}G^{\mu\nu}G_{\mu\nu}$, ${\alpha_s\over\pi}{\cal S}[G^{\rho}_{\mu}G_{\rho\nu}]$, and quark operators, ${\cal S}[\bar q\gamma_\mu iD_\nu q]$, $\bar q{\cal S}(\gamma_\mu iD_\nu)\gamma_5 q$.
The nucleon matrix elements of the gluon operators are spin-independent and have already been discussed in Refs.~7) and~13).
The matrix element of the quark operator, $\langle{\cal S}[\bar q\gamma_\mu iD_\nu q]\rangle_N$, is also spin-independent and is related to the unpolarized quark distribution $f_1$,~\cite{rf:HL} while $\langle\bar q{\cal S}(\gamma_\mu iD_\nu)\gamma_5 q\rangle_N$
is spin-dependent and
\begin{eqnarray*}
     \langle\bar q{\cal S}(\gamma_\mu iD_\nu)\gamma_5 q\rangle_N
 &=& a_1s_{\{\mu} \hat p_{\nu\}}.
\end{eqnarray*}
By neglecting the operator including the quark mass, $a_1$ can be related to the first moment of the longitudinal quark-spin distribution, $g_2$, and it is given at tree-level by~\cite{rf:KYU}
\begin{eqnarray*}
 a_1=-2\int_0^1dxxg_2(x).
\end{eqnarray*}

Very recently, measurements of $g_2$ for the proton and the neutron have begun.~\cite{rf:g2p}~\cite{rf:g2n}
We have calculated $a_1$ for the proton using the data for $g_2$ in Ref.~16) over the range $0.075<x<0.8$ and $1.3<Q^2<10\,({\rm(GeV}/c)^2$ for the proton and those in Ref.~17) over the range $0.06<x<0.70$ and $1.0<Q^2<17.0\,({\rm(GeV}/c)^2$ for the neutron with the results
\begin{eqnarray*}
     a^u_1 = 0.05 \pm 0.04,\qquad
     a^d_1 =-0.08 \pm 0.13.
\end{eqnarray*}

In addition to the above dimension-three and dimension-four operators we take into account the dimension-six four-quark operators, $\bar qq\bar qq$, $\bar qq\bar q\gamma_\mu q$, $\bar qq\bar q\gamma_\mu\gamma_5 q$ and $\bar qq\bar q\gamma_5\sigma_{\mu\nu} q$, since four-quark operators are known to give the largest contribution among higher order operators in the QCD sum rule for the nucleon in the vacuum.~\cite{rf:RRY}~\cite{rf:HL}
In the vacuum, the matrix elements of the four-quark operators are estimated by the factorization hypothesis;~{\cite{rf:SVZ}\cite{rf:RRY}} i.e. it is assumed that the vacuum contribution dominates in the intermediate states: 
$\langle{\cal O}_1{\cal O}_2\rangle_0\approx\langle{\cal O}_1\rangle_0\langle{\cal O}_2\rangle_0$.
Similarly, for the nucleon matrix element, we assume that the contribution from the one-nucleon state dominates in the intermediate states:
\begin{eqnarray*}
  \langle{\cal O}_1{\cal O}_2\rangle_N
 &\equiv&\langle N|{\cal O}_1{\cal O}_2|N \rangle-\langle N|N \rangle\langle{\cal O}_1{\cal O}_2\rangle_0\cr
 &\approx&{\langle N|{\cal O}_1|N \rangle\langle N|{\cal O}_2|N \rangle
     \over\langle N|N \rangle}
    -\langle{\cal O}_1\rangle_0\langle{\cal O}_2\rangle_0\langle N|N \rangle\cr
 &=&\langle{\cal O}_1\rangle_N\langle{\cal O}_2\rangle_0
    +\langle{\cal O}_1\rangle_0\langle{\cal O}_2\rangle_N .
\end{eqnarray*}
Thus we assume $\langle\bar qq\bar qq\rangle_N=2\langle\bar qq\rangle_0\langle\bar qq\rangle_N$, $\langle\bar qq\bar q\gamma_\mu q\rangle_N=\langle\bar qq\rangle_0\langle\bar q\gamma_\mu q\rangle_N$, $\langle\bar qq\bar q\gamma_\mu\gamma_5 q\rangle_N=\langle\bar qq\rangle_0\langle\bar q\gamma_\mu\gamma_5 q\rangle_N$ and $\langle\bar qq\bar q\gamma_5\sigma_{\mu\nu} q\rangle_N=\langle\bar qq\rangle_0\langle\bar q\gamma_5\sigma_{\mu\nu} q\rangle_N$.

For completeness we list here the values which were used in the calculation. 
The spin-independent proton matrix elements of the operators are 
\begin{eqnarray*}
  &&\langle u^\dagger u\rangle_p=2,\qquad\langle d^\dagger d\rangle_p=1,\qquad
    \langle\bar u u\rangle_p=3.46,\qquad \langle\bar d d\rangle_p=2.96, \cr
  &&i\langle{\cal S}[\bar u\gamma_\mu D_\nu u]\rangle_p=222\;{\rm MeV},\qquad
    i\langle{\cal S}[\bar d\gamma_\mu D_\nu d]\rangle_p=95\;{\rm MeV},\cr 
  &&\langle{\alpha_s\over\pi}G_{\mu\nu}G^{\mu\nu}\rangle_p =-738 \;{\rm MeV},\qquad
   \langle{\alpha_s\over\pi}{\cal S}[G_{\mu 0}G^{\mu 0}]\rangle_p =-50 \;{\rm MeV}.
\end{eqnarray*}

The condensates of the operators in the vacuum are
$\langle\bar uu\rangle_0 =\langle\bar dd\rangle_0 = -(250\;{\rm MeV})^3$ and
$\langle{\alpha_s\over\pi}G^2\rangle_0 = (330\;{\rm MeV})^4$.~\cite{rf:RRY}
In the QCD sum rule for the nucleon in the vacuum, $\w$ is determined to be $2.2\ {\rm GeV}$ by Borel stability analysis.

$\wp$ and $\wm$ are determined by Borel stability analysis.
Namely, we search for the values of  $\wp$ and $\wm$ for which the calculated
strength $\alpha$ has the most stable plateau as a function of the Borel mass $\MB$ both in the triplet and singlet channels.
We find that the optimum choice is  $\wp\approx 1.3\ {\rm GeV}$ and $\wm\approx 1.3\ {\rm GeV}$.
Figures~1 and 2 display how sensitive the Borel stability is to $\wp$ and $\wm$.
Figure~1 displays $\alpha$ vs. the Borel mass for $\wp = 1.3\ {\rm GeV}$ and $\wm =1.2\ {\rm GeV}, 1.3\ {\rm GeV}$ and $1.4\ {\rm GeV}$.
Figure~2 displays the same for $\wm = 1.3\ {\rm GeV}$ and $\wp =1.1\ {\rm GeV}, 1.3\ {\rm GeV}$ and $1.5\ {\rm GeV}$.
One sees that the Borel stability is much more sensitive to $\wm$ than to $\wp$.
This is because the $\wm$ dependence of the coefficients of the dimension-three and dimension-four operators is much stronger than the $\wp$ dependence.

\begin{figure}
  \begin{center}
    \leavevmode
    \psfig{figure=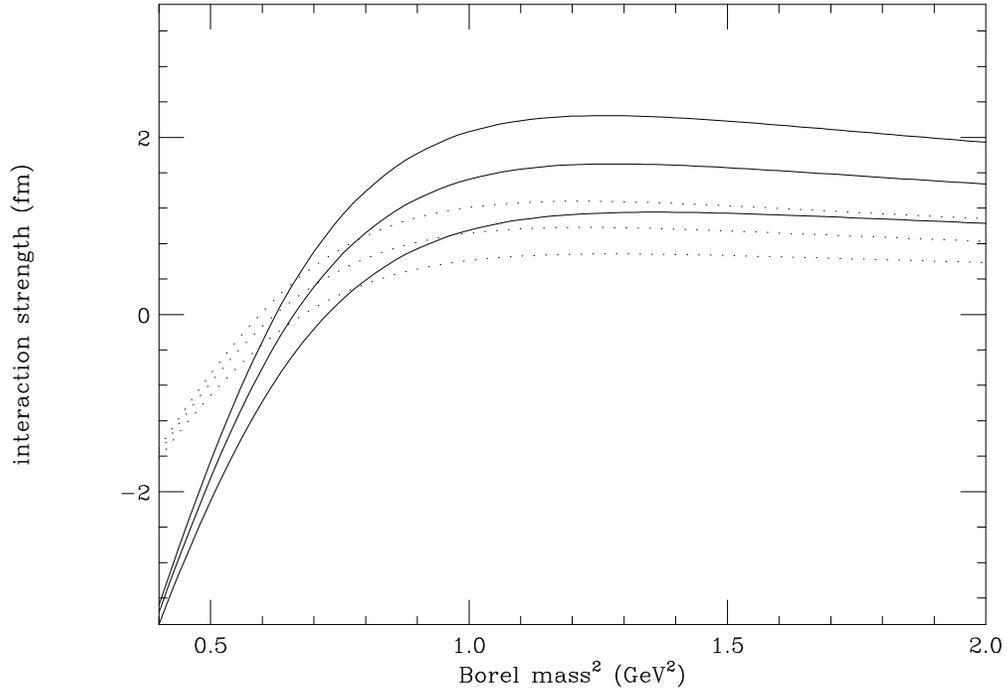,angle=90,width=5.2in}
  \end{center}
  \caption{The calculated strength $\alpha$ as a function of the Borel mass squared, $\MBS$. The solid lines represent the spin triplet channel, and the dotted lines represent the spin singlet channel by fixing $\wp$ to $1.3\ {\rm GeV}$. The top, middle and bottom lines are for $\wm=1.2\ {\rm GeV}$, $\wm=1.3\ {\rm GeV}$ and $\wm=1.4\ {\rm GeV}$, respectively.}
  \label{fig:1}
\end{figure}
\begin{figure}
  \begin{center}
    \leavevmode
    \psfig{figure=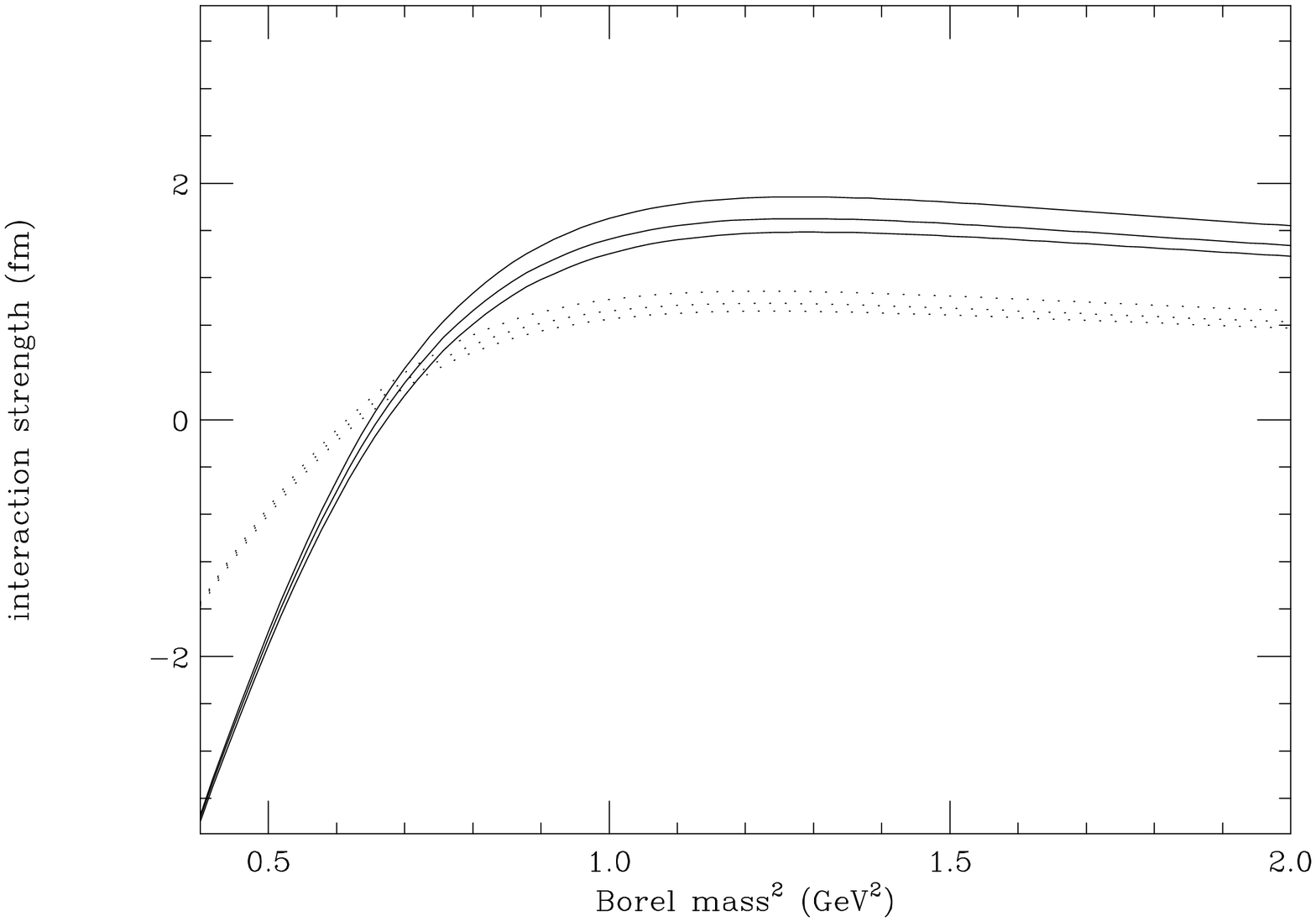,angle=90,width=5.2in}
  \end{center}
  \caption{The calculated strength $\alpha$ as a function of the Borel mass squared, $\MBS$. The solid lines corresponds to the spin triplet channel, and the dotted lines corresponds to the spin singlet channel by fixing $\wm$ to $1.3\ {\rm GeV}$. The top, middle and bottom lines are for $\wp=1.1\ {\rm GeV}$, $\wp=1.3\ {\rm GeV}$ and $\wp=1.5\ {\rm GeV}$, respectively.}
  \label{fig:2}
\end{figure}

The $\alpha$ is determined by taking the maximum values for $\wp= 1.3\ {\rm GeV}$ and $\wm= 1.3\ {\rm GeV}$ as
\begin{eqnarray*}
 \alpha_{pn}^{3}&=&1.7\;{\rm fm},\cr
 \alpha_{pn}^{1}&=&1.0\;{\rm fm}.\cr
\end{eqnarray*}
We see that the spin-dependent part is rather smaller than the spin-independent part.
This is due to the small matrix elements of the spin-dependent operators.
From Figs.~1 and 2 we should expect errors of about 30\% due to the choice of $\wp$ and $\wm$.
The errors due to the uncertainties of the nucleon matrix elements of the dimension-three and dimension-four operators are about 10\% and 20\%, respectively.
Combining all the errors together, the above results have errors of approximately 40\%.

Let us compare the above results with experimental facts.
Experimentally, the scattering length and the effective range have been found to be
\begin{eqnarray*}
 a_{pn}^{3}&=&-5.39\;{\rm fm},\qquad r_{pn}^{3}=1.75\;{\rm fm},\cr
 a_{pn}^{1}&=&23.7\;{\rm fm},\qquad r_{pn}^{1}=2.73\;{\rm fm}.\cr
\end{eqnarray*}

In the spin-singlet channel, the scattering length is so large that the strength, $\alpha_{pn}^{1}$, is expected to be approximated well by ${r_{pn}^{1}}/2$,
\begin{eqnarray*}
\alpha_{pn}^{1} \approx {r_{pn}^{1}}/2 = 1.37\;{\rm fm} ,
\end{eqnarray*}
which is in rather good agreement with the calculated result.

In the spin-triplet channel, the scattering length is not small. However, it is not as large as in the spin-singlet channel.
Therefore, it is not easy to estimate the strength $\alpha_{pn}^{3}$ from experimental observables.
However, there is a loosely bound state, deuteron, in the spin-triplet channel, while there is an almost bound state in the spin-singlet channel.
This implies that the interaction in the spin-triplet channel is stronger (but not very stronger) than that in the spin-singlet channel.
This tendency is consistent with the calculated results.

\section{Summary and outlook}
In this paper we have studied spin-dependent nucleon-nucleon ($NN$) interactions in the QCD sum rules.
The basic object of our study is the spin-dependent nucleon correlation function, whose matrix element is taken with respect to the one-nucleon state.

The dispersion integral of the correlation function around the nucleon threshold has been investigated in detail.
The integral is given by the sum of the threshold contribution, which is due to the second-order pole term proportional to the scattering length, the continuum contribution, and the bound state contribution. 
When the interaction is weak, the integral is dominated by the threshold contribution and is proportional to the scattering length.
As the interaction becomes stronger, the continuum contribution also becomes important.
When the interaction is just so strong as to form a bound state, both the threshold contribution and the continuum contribution diverge, but their sum is finite.
Thus the sum of these two contributions is given by the same form as in the case of the weak interaction.
However, in this case the scattering length is replaced by one half of the effective range.  
Based on this observation 
we have defined the $NN$ interaction strength through the dispersion integral around the nucleon threshold.

In the OPE of the correlation function, new operators, such as $\bar q\gamma_\mu\gamma_5q$, $\bar q\gamma_5\sigma_{\mu\nu}q$, have to be taken into account.
These operators do not vanish when the matrix element is taken with respect to the spin-nonaveraged one-nucleon state.
We have calculated the Wilson coefficients of such operators.

The obtained sum rules relate the spin-dependent $NN$ interaction strengths with the spin-dependent nucleon matrix elements of the quark-gluon composite operators.
The spin-dependent nucleon matrix elements such as
$\langle\bar q\gamma_\mu\gamma_5 q\rangle_N$ and $\langle\bar q\gamma_5\sigma_{\mu\nu} q\rangle_N$ are related with the spin-dependent structure functions of the nucleon, $g_1$, $g_2$ and $h_1$.
We found that the interaction strength in the spin-singlet channel is weaker than in the spin-triplet channel, but that the spin-dependent part of the interaction strength is considerably smaller than the spin-independent part.
Experimentally, it has been found that there is a loosely bound state, deuteron, in the spin-triplet channel, while there is an almost bound state in the spin-singlet channel, which implies that the interaction is slightly stronger in the spin-triplet channel than in the spin-singlet channel.
This seems to be consistent with the sum rule result.
In the spin-singlet channel, the scattering length is so large that the interaction strength can be estimated by using the observed effective range, while in the spin-triplet channel the absolute value of the interaction strength is difficult to obtain from observables.
The empirical interaction strength thus obtained in the spin-singlet channel agrees rather well with the sum rule calculation. 

The method used in the present paper can be extended to other hadron-nucleon channels.
In particular, it is straightforward to apply it to the hyperon-nucleon channels.
The obtained sum rules would relate the hyperon-nucleon interaction strengths with the nucleon matrix elements of the quark-gluon operators which include strange quark operators in addition to up and down quark operators.
Also, the present sum rules can be used in the opposite way.
Namely, if one knows detailed information on various spin-dependent hadron-nucleon interaction strengths, one can obtain spin-dependent matrix elements of quark-gluon operators with respect to the one-nucleon state. This provides us with information such as the spin content of the nucleon.

\vskip 14pt
\begin{center}
{\bf Acknowledgements}
\end{center}
\vskip 14pt

We would like to thank Professor K.~Yazaki for valuable discussions and Professor M.~Oka for bringing our attention to the spin-dependent correlation function. We are also grateful to Professor H. Terazawa for careful reading of the manuscript.

\end{document}